\documentclass[aps, preprint, groupedaddress, floatfix]{revtex4}

\usepackage[english]{babel}

\usepackage[letterpaper,top=2cm,bottom=2cm,left=3cm,right=3cm,marginparwidth=1.75cm]{geometry}

\usepackage{amsmath}
\usepackage{graphicx}
\usepackage[colorlinks=true, allcolors=blue]{hyperref}
\usepackage[toc,page,titletoc]{appendix}

\begin{document}

\title{Dynamical structural instability and its
implication on the physical properties of infinite-layer nickelates}

\author{Chengliang Xia$^{1,2}$, Jiaxuan Wu$^2$, Yue Chen$^1$ and Hanghui Chen$^{2,3}$}
\affiliation{$^1$Department of Mechanical Engineering, 
The University of Hong Kong, Pokfulam Road, Hong Kong SAR, China\\ 
$^2$NYU-ECNU Institute of Physics, NYU Shanghai, Shanghai 200122, China\\
$^3$Department of Physics, New York University, New York, NY 10003, USA}

\begin{abstract}
We use first-principles calculations to find that 
in infinite-layer nickelates $R$NiO$_2$, the widely 
studied tetragonal $P4/mmm$ structure is only 
dynamically stable for early lanthanide elements $R$ = La-Sm.
For late lanthanide elements $R$ = Eu-Lu, an imaginary 
phonon frequency appears at $A=(\pi,\pi,\pi)$ point. 
For those infinite-layer nickelates,
condensation of this phonon mode into the $P4/mmm$ structure 
leads to a more energetically favorable $I4/mcm$ structure that is 
characterized by an out-of-phase rotation of 
``NiO$_4$ square". Special attention is 
given to two borderline cases: PmNiO$_2$ and SmNiO$_2$, 
in which both the $P4/mmm$ structure and the $I4/mcm$
structure are local minimums, and the energy difference 
between the two structures can be fine-tuned by epitaxial 
strain. Compared to the $P4/mmm$ structure, $R$NiO$_2$ 
in the $I4/mcm$ structure has a substantially reduced 
Ni $d_{x^2-y^2}$ bandwidth, a smaller Ni $d$ 
occupancy, a ``cleaner" Fermi surface with a 
lanthanide-$d$-derived electron pocket suppressed at 
$\Gamma$ point, and a decreased critical 
$U_{\textrm{Ni}}$ to stabilize long-range antiferromagnetic 
ordering. All these features imply enhanced correlation effects 
and favor Mott physics. Our work reveals the importance of structure-property 
relation in infinite-layer nickelates, in particular
the spontaneous ``NiO$_4$ square" rotation 
provides a tuning knob to render $R$NiO$_2$ in the $I4/mcm$ 
structure a closer analogy to superconducting infinite-layer cuprates.
\end{abstract}

\maketitle

\section{Introduction}

The discovery of superconductivity in infinite-layer 
nickelates Sr$_{x}$Nd$_{1-x}$NiO$_2$~\cite{Li2019} has drawn 
great attention~\cite{Hepting2020,Zeng2020,Li2020a,Gu2020a,Goodge2021,Wang2021,Zhao2021,Lu2021} 
because the parent material NdNiO$_2$ 
has similar crystal and 
electronic structures to those of infinite-layer 
cuprate CaCuO$_2$~\cite{Sawatzky2019}, which exhibits high-temperature 
unconventional superconductivity upon 
doping~\cite{Smith1991,Azuma1992}. 
Both NdNiO$_2$ and CaCuO$_2$ crystallize 
in a simple tetragonal $P4/mmm$ structure in which 
Ni (Cu) and O atoms form a flat 
``NiO$_4$ (CuO$_4$) square"~\cite{Siegrist1988,Li2020, Wang2020c}. 
The $P4/mmm$ crystal structure has only two degrees of freedom:
lattice constants $a$ and $c$. As for the  
non-interacting electronic structure, 
first-principles calculations show that CaCuO$_2$ has only 
one Cu $d_{x^2-y^2}$-derived band that crosses the 
Fermi level, while NdNiO$_2$ has two bands crossing the
Fermi level~\cite{Jiang2019,Botana2020,Karp2020a,Adhikary2020,Zhang2020d}. One is Ni $d_{x^2-y^2}$-derived band
and the other band is derived from Nd $d$ orbitals and 
an interstitial $s$ orbital~\cite{Nomura2019,Gu2020,Hirayama2020}. 
So far, the minimum 
theoretical model that is adequate to describe the
low-energy physics of infinite-layer NdNiO$_2$ has been
under intensive debate and several different
mechanisms for superconductivity in
Sr$_{x}$Nd$_{1-x}$NiO$_2$ have been proposed~\cite{Hu2019, Wang2020,Jiang2020,Si2020,Geisler2020,Sakakibara2020,He2020,Wu2020,Werner2020,Zhang2020,Wang2020b,Zhang2020c,Bernardini2020,Bernardini2020a,Liu2021,Wan2021,Plienbumrung2021,Malyi2021,Peng2021,Choubey2021,Kang2021}.
Albeit there are many important differences, 
one thing in common is that 
all first-principles calculations use
the $P4/mmm$ crystal structure of NdNiO$_2$ (either
experimental one or theoretical optimized one), 
based on which one-particle band structure calculations 
(using density functional theory and its Hubbard $U$ extension)~\cite{Krishna2020,
Choi2020,Liu2020a,Choi2020a,Zhang2020a,Zhang2021} or
more sophisticated many-body electronic structure
calculations (such as dynamical mean field theory 
and GW)~\cite{Katukuri2020,Olevano2020,
Karp2020,Ryee2020,Lechermann2020,Lechermann2020a,
Leonov2020a,Petocchi2020,Lechermann2021,
Kitatani2020,Karp2021,Leonov2021,Kutepov2021} 
are performed.
On the experimental side, in addition to
Sr$_{x}$Nd$_{1-x}$NiO$_2$, recently superconductivity 
is also observed in 
Sr$_{x}$Pr$_{1-x}$NiO$_2$~\cite{Osada2020,Osada2020a,Ren2021},
Sr$_{x}$La$_{1-x}$NiO$_2$~\cite{Osada2021} and  
Ca$_{x}$La$_{1-x}$NiO$_2$~\cite{Zeng2021,Puphal2021}. 
Thus it is anticipated that
superconductivity should be observed in the entire 
lanthanide series of infinite-layer nickelates
$R$NiO$_2$. In particular, 
Refs.~\cite{Kapeghian2020,Been2021} perform a systematic 
study on the electronic structure of $R$NiO$_2$ 
in the $P4/mmm$ structure as 
$R$ traverses the lanthanide series and find promising 
trends that favor superconductivity.

In this work, we use first-principles calculations to 
show that the widely studied $P4/mmm$ structure of 
infinite-layer nickelates is only dynamically stable 
for early lanthanide elements $R$ = La-Sm.
For late lanthanide elements $R$ = Eu-Lu,
an imaginary phonon mode appears at $A=(\pi,\pi,\pi)$ point 
in the $P4/mmm$ structure of $R$NiO$_2$. The imaginary 
phonon mode corresponds to an out-of-phase rotation 
of ``NiO$_4$ square" about the $z$ axis. Condensation of 
this unstable phonon mode into the $P4/mmm$ structure 
leads to a more energetically favorable crystal structure 
with lower symmetry (space group $I4/mcm$). Attention
is given to two borderline 
cases PmNiO$_2$ and SmNiO$_2$, which have two 
local minimums: the $P4/mmm$ structure and 
the $I4/mcm$ structure.
Epitaxial strain can be used to fine-tune the energy 
difference between the two crystal structures.

Compared to the $P4/mmm$ structure,  
infinite-layer $R$NiO$_2$ in the new $I4/mcm$ structure has a 
distinct electronic structure: 
the Ni $d_{x^2-y^2}$ bandwidth is substantially reduced 
(by about 0.5 eV) and Ni $d$ occupancy 
decreases; the Fermi 
surface becomes ``cleaner" because one lanthanide-$d$-derived 
electron pocket disappears at the $\Gamma$ point;
the critical $U_{\textrm{Ni}}$ to stabilize a long-range antiferromagnetic 
ordering in $R$NiO$_2$ is smaller. 
All these features imply
that with the ``NiO$_4$ square" rotation, correlation effects 
will be enhanced and Mott physics will 
play a more prominent role in the new $I4/mcm$ structure than 
in the $P4/mmm$ structure, when local interaction is added on 
Ni $d$ orbitals~\cite{Lee2006}. 
In particular, our results suggest that among the 
lanthanide series of 
infinite-layer nickelates, SmNiO$_2$ is the most 
promising candidate to crystallize in the 
$I4/mcm$ structure, which renders it a closer analogy to  
superconducting infinite-layer cuprates. 

\section{Computational Details}
We perform density functional theory
(DFT)~\cite{Hohenberg1964,KohnW.andSham1965} calculations within
the~\textit{ab initio} plane-wave approach, as implemented in the
Vienna Ab-initio Simulation Package
(VASP)~\cite{Payne1992,Kresse1996}. We use projected augmented wave
(PAW) pseudopotentials with the 4$f$ electrons placed in the core
(except for La), explicitly to avoid complication that arises from
treating the localized 4$f$ electrons.  We employ generalized gradient
approximation (GGA) for the exchange-correlation functional with
Perdew-Burke-Ernzerhof (PBE) parametrization~\cite{Perdew1996}.  The
theoretical lattice constants of $R$NiO$_2$ are in good agreement with
the available experimental structure information (see Supplementary
Materials~\cite{SI} Sec.~I) ~\cite{Wang2020c,Zhang2021,Lin2021arxiv}.
We use an energy cutoff of 600 eV.  Charge self-consistent
calculations are converged to 10$^{-7}$ eV. Both cell and internal
atomic positions are fully relaxed until each force component is
smaller than 1 meV/\AA~and pressure on the cell is smaller than 0.1
kB.  We use the finite-displacement method to calculate the full
phonon dispersion with the aid of Phonopy~\cite{Togo2015}. A supercell
that consists of $3\times 3 \times 3$ primitive cells is used to
calculate the force constants and dynamical matrices.  The primitive
cell of the $P4/mmm$ structure has one formula of $R$NiO$_2$ (i.e. 4
atoms), while the primitive cell of the $I4/mcm$ structure has two
formulae of $R$NiO$_2$ (i.e. 8 atoms) in order to accommodate the
rotation of ``NiO$_4$ square".  For the 4-atom $P4/mmm$ cell, we use a
Monkhorst-Pack \textbf{k} mesh of $14\times 14 \times 14$ to sample
the first Brillouin zone. For the 8-atom $I4/mcm$ cell, we use a
Monkhorst-Pack \textbf{k} mesh of $10\times 10 \times 10$ to sample
the first Brillouin zone. For ease of comparison, when calculating the
electronic structure, Fermi surface and long-range magnetic ordering,
we use the 8-atom cell for both $P4/mmm$ and $I4/mcm$ structures. This
cell-doubling is also necessary to accommodate the rocksalt
antiferromagnetic ordering in the $P4/mmm$ structure. To
  study the energy evolution as a function of ``NiO$_4$ square''
  rotation, we use a linear interpolation and generate an intermediate
  crystal structure between the fully-relaxed $P4/mmm$ and $I4/mcm$
  crystal structures. The ``NiO$_4$ square'' rotation angle
  continuously changes with an interpolation parameter $\lambda$~(more
  details are found in Supplementary Materials~\cite{SI}
  Sec.~II). When we impose a bi-axial strain on infinite-layer
  nickelates, the out-of-plane $c$ axis is fully relaxed in order to
  minimize the total energy.  When calculating
Sr$_{x}R_{1-x}$NiO$_2$, we use virtual crystal approximation
(VCA)~\cite{Bellaiche2000}.  To break spin symmetry and study magnetic
order, we use the charge-only DFT+$U$+$J$
method~\cite{Park2015,Chen2015,Chen2016} (by setting LDAUTYPE = 4 in
VASP). This method is such that the exchange-correlation functional
only depends on charge density but not on spin density; thereby, spin
symmetry is only broken by the $U/J$ terms that are added to the
Kohn-Sham potential, while the exchange splitting that arises from the
spin-dependent exchange-correlation functional is disabled. By setting
the parameter $U/J = 0$ in the charge-only DFT+$U$+$J$ method, the
non-spin-polarized DFT results are recovered.  The calculations of Ni
$d$ projected magnetic moment and Ni $d$ occupancy use the default
VASP value for the radius of sphere which is 1.11~\AA. We also use the
charge-only DFT+$U$+$J$ method to test phonon spectrum for a few
representative
$R$NiO$_2$ (see Supplementary Materials~\cite{SI} Sec.~VIII)
~\cite{Lin2021arxiv,Hayward1999,Hayward2003}.

\section{Results and Discussion}

\subsection{Phonon spectrum}

\begin{figure}[t]
\includegraphics[width=\textwidth]{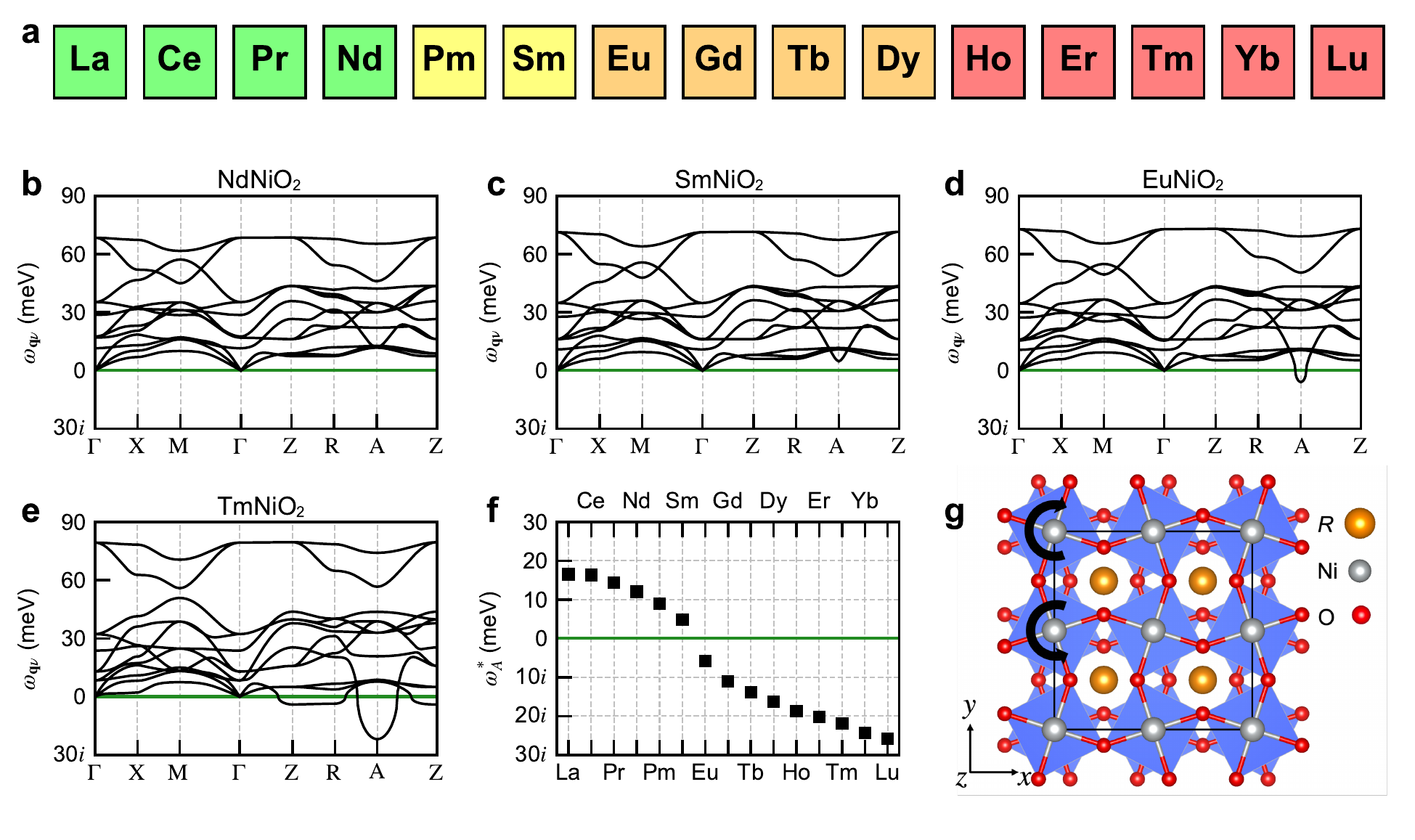}
\caption{\label{fig:phonon} Phonon properties of 
infinite-layer $R$NiO$_2$ in the $P4/mmm$ structure. 
\textbf{a}: The series of lanthanide 
elements. Different colors are used to distinguish 
four categories of phonon dispersions, whose prototypes are shown 
in \textbf{b}-\textbf{e}. \textbf{b}: Phonon dispersion of 
infinite-layer NdNiO$_2$ in the $P4/mmm$ structure.  
\textbf{c}: Phonon dispersion of infinite-layer SmNiO$_2$ 
in the $P4/mmm$ structure.  \textbf{d}: Phonon dispersion 
of infinite-layer EuNiO$_2$ in the $P4/mmm$ structure.  
\textbf{e}: Phonon dispersion of infinite-layer TmNiO$_2$ 
in the $P4/mmm$ structure. \textbf{f}: Frequency of the 
lowest phonon mode at $A=(\pi,\pi,\pi)$ point
$\omega^*_{A}$ for the entire 
lanthanide series of infinite-layer $R$NiO$_2$ in the $P4/mmm$ 
structure.
\textbf{g}: The lowest phonon mode at $A$ point of 
infinite-layer $R$NiO$_2$ in the $P4/mmm$ structure. }
\end{figure}

We calculate the phonon dispersion of the fully-relaxed
$P4/mmm$ structure for the entire lanthanide series of 
infinite-layer $R$NiO$_2$. We find
that the complete set of phonon dispersions
(see Supplementary Materials~\cite{SI} Sec.~II)  
can be classified into
four categories, which we use colors to distinguish in panel 
\textbf{a}. The first category includes $R$=La-Nd 
(denoted by green). 
NdNiO$_2$ is the prototype, whose full phonon dispersion 
is shown in panel \textbf{b}.
In this category, the full phonon dispersion is free of 
imaginary modes and the $P4/mmm$ crystal 
structure is dynamically stable.
The second category includes $R$=Pm, Sm (denoted by yellow).
SmNiO$_2$ is the prototype, whose full phonon dispersion 
is shown in panel \textbf{c}.
In this category, a soft phonon develops at 
$A=(\pi,\pi,\pi)$ point, 
implying a potentially unstable mode.
The third category includes $R$=Eu-Dy (denoted by orange).
EuNiO$_2$ is the prototype, whose full phonon dispersion 
is shown in panel \textbf{d}.
In this category, the frequency of the lowest phonon mode at 
$A$ point (marked as $\omega^*_{A}$)
becomes imaginary and 
the $P4/mmm$ crystal 
structure is dynamically unstable.
The last category includes $R$=Ho-Lu (denoted by red).
TmNiO$_2$ is the prototype, whose full phonon dispersion 
is shown in panel \textbf{e}. 
In this category, the lowest phonon modes at 
multiple \textbf{q} points become imaginary in the 
phonon dispersion, indicating that the $P4/mmm$ crystal 
structure is far from stable. In panel \textbf{f}, 
we compare the frequency
of the lowest phonon mode at $A$ point $\omega^*_{A}$ for the
entire lanthanide series. We find that 
$\omega^*_{A}$ monotonically decreases from La to Lu and 
becomes imaginary when $R$=Eu and beyond.
In panel \textbf{g}, we  
show the lowest phonon mode at $A$ point of the 
$P4/mmm$ structure, which is an out-of-phase rotation 
of ``NiO$_4$ square" about the $z$ axis. 
Infinite-layer nickelates with early lanthanide elements such 
as NdNiO$_2$ are stable against this ``NiO$_4$ square" 
rotation, and their equilibrium structure is
the widely-studied $P4/mmm$ structure. However, 
infinite-layer nickelates with late lanthanide elements 
become dynamically unstable when the ``NiO$_4$ square" 
rotates. Condensation of this unstable mode into the 
$P4/mmm$ structure will lead to 
a more energetically favorable crystal structure, which 
in turn results in a new electronic structure.

\subsection{Rotation of ``NiO$_4$ square" and the 
new $I4/mcm$ structure}

\begin{figure}[t]
\includegraphics[width=0.85\textwidth]{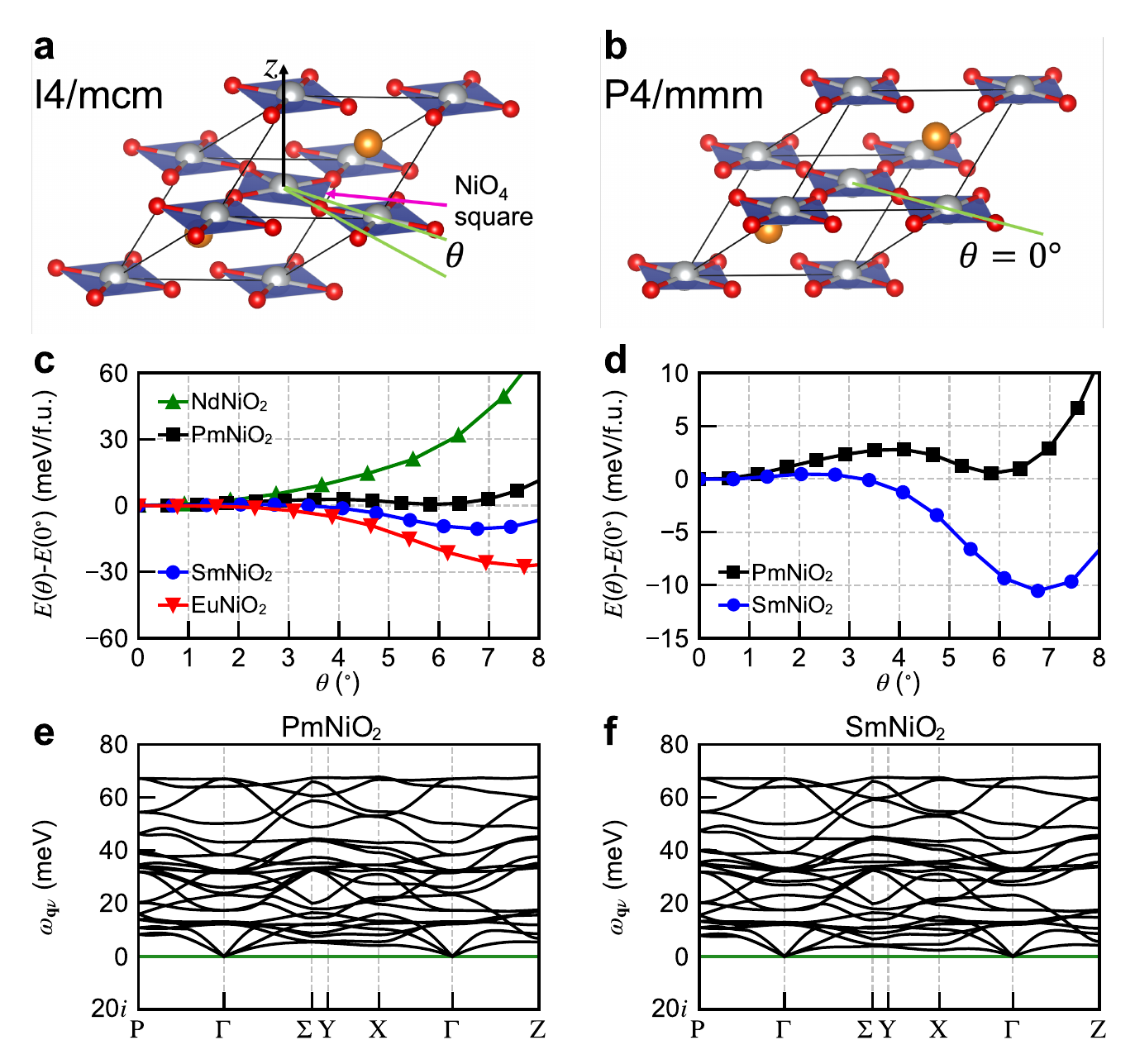}
\caption{\label{fig:local-minimum} \textbf{a}: The 
primitive cell of infinite-layer nickelate 
$R$NiO$_2$ in the $I4/mcm$ structure. 
$\theta$ is the rotation angle of ``NiO$_4$ square" 
along the $z$ axis. \textbf{b}: A doubled unit cell 
of infinite-layer nickelate
$R$NiO$_2$ in the $P4/mmm$ structure (the 
$P4/mmm$ primitive cell is doubled along 
[111] direction). $\theta=0^{\circ}$ 
in the $P4/mmm$ structure.
\textbf{c}: The energy evolution of $R$NiO$_2$ 
($R$ = Nd, Pm, Sm and Eu) as a 
function of the ``NiO$_4$ square" rotation 
angle $\theta$. The $P4/mmm$ 
structure is used as the energy reference. 
\textbf{d}: The zoom-in view of 
the energy evolution of PmNiO$_2$ and 
SmNiO$_2$ as a function 
of the ``NiO$_4$ square" rotation angle $\theta$. 
\textbf{e} and \textbf{f}: 
The phonon dispersions of infinite-layer PmNiO$_2$ 
and SmNiO$_2$ in the $I4/mcm$ 
structure, respectively.}
\end{figure}

When the lowest phonon frequency at the $A$ point in the $P4/mmm$ 
structure of $R$NiO$_2$ becomes imaginary ($R$ = Eu-Lu), it means 
that condensation of this unstable phonon mode 
into the $P4/mmm$ structure 
can decrease the total energy and will result in a new crystal 
structure with lower symmetry. Such a crystal structure is shown 
in panel \textbf{a} of Fig.~\ref{fig:local-minimum}, 
which has space group $I4/mcm$. 
The primitive cell of the $I4/mcm$ structure has 8 atoms, which 
has three degrees of freedom: in addition to 
the two lattice constants $a$ and $c$, there is an angle $\theta$ 
that characterizes the out-of-phase rotation of ``NiO$_4$ square" about 
the $z$ axis. 
When $\theta = 0$, the $I4/mcm$ structure is reduced to 
the $P4/mmm$ structure. For ease of comparison with the 
$I4/mcm$ structure, we show the 
two-Ni unit cell of the $P4/mmm$ structure (doubling 
the 4-atom primitive cell along the $[111]$ direction) 
in panel \textbf{b}. For subsequent electronic structure, Fermi 
surface and long-range magnetic ordering calculations of 
infinite-layer $R$NiO$_2$, we use the 8-atom cell for both 
the $I4/mcm$ structure and the $P4/mmm$ structure. 

To get a better understanding of the new $I4/mcm$ structure, 
we calculate the energy evolution of $R$NiO$_2$ as a function 
of the rotation angle $\theta$ (see panel \textbf{c} of 
Fig.~\ref{fig:local-minimum}, the calculation
  details can be found in Supplementary Materials Sec. II~\cite{SI}).
We select $R$ = Nd, Pm, Sm and Eu, 
which are near the phase boundary where the lowest 
phonon frequency at $A$ point becomes 
imaginary. We find that the total energy of NdNiO$_2$ 
monotonically increases with the rotation angle $\theta$, 
indicating that the $P4/mmm$ structure (i.e. $\theta = 0$) 
is stable against the rotation of ``NiO$_4$ square".
By contrast, the total energy of EuNiO$_2$ first 
decreases with the rotation angle $\theta$ and then 
increases. The energy minimum is at $\theta = 7.7^{\circ}$. 
This clearly shows that the $P4/mmm$ structure is not 
dynamically stable in infinite-layer EuNiO$_2$. 
PmNiO$_2$ and SmNiO$_2$ (the second category) 
exhibit more interesting features in that they have two local 
minimums: one is at $\theta = 0$ ($P4/mmm$ structure) and the 
other is at $\theta > 0$ ($I4/mcm$ structure), as is shown in 
panel \textbf{d}. For PmNiO$_2$, the energy of the $P4/mmm$ 
structure is slightly lower than that of the 
$I4/mcm$ structure by 0.5 meV/f.u.  
For SmNiO$_2$, the energy order is reversed and 
the $I4/mcm$ structure becomes more stable 
than the $P4/mmm$ structure by 10.5 meV/f.u. Furthermore,
we calculate the energy barrier from the $P4/mmm$ 
structure to the $I4/mcm$ structure. We find that 
the barrier decreases from 2.8 meV/f.u. for PmNiO$_2$ to 0.5 meV/f.u.
for SmNiO$_2$. Next we test that after condensation of the 
``NiO$_4$ square" rotation mode, the $I4/mcm$ structure becomes
dynamically stable in some infinite-layer nickelates. We 
perform the phonon calculation of the $I4/mcm$ structure 
for PmNiO$_2$ and SmNiO$_2$. The phonon dispersions are 
shown in panels \textbf{e} and \textbf{f}.
We find that the phonon dispersion of the $I4/mcm$ 
structure is free from imaginary frequencies for PmNiO$_2$ 
and SmNiO$_2$. We make two comments here. First, 
while SmNiO$_2$ has two local 
minimums, considering the facts that 1) 
its $I4/mcm$ structure is energetically more 
favorable than the $P4/mmm$ structure, 2) the energy barrier 
for SmNiO$_2$ to transition from the $P4/mmm$ structure 
to the $I4/mcm$ structure 
is tiny (0.5 meV/f.u.), and 3) 
its $I4/mcm$ structure is dynamically 
stable, we argue that in experiments SmNiO$_2$ is most likely stabilized in 
the $I4/mcm$ structure. Second, the ``NiO$_4$ square" 
rotation is the first structural 
distortion that will appear in the 
$P4/mmm$ structure of infinite-layer nickelates 
$R$NiO$_2$ when the lanthanide element $R$ 
traverses from La to Lu.
For late lanthanide elements (such as Ho-Lu), 
more complicated structural distortions are expected to 
emerge in infinite-layer $R$NiO$_2$. The purpose of 
the current study is to show that 
just by including one more degree of freedom in 
the crystal structure of 
$R$NiO$_2$ (i.e. ``NiO$_4$ square" rotation), 
the resulting electronic 
structure trends can be qualitatively different from those of the 
$P4/mmm$ structure (see discussion below). 

\subsection{Epitaxial strain}

\begin{figure}[t]
\includegraphics[width=0.8\textwidth]{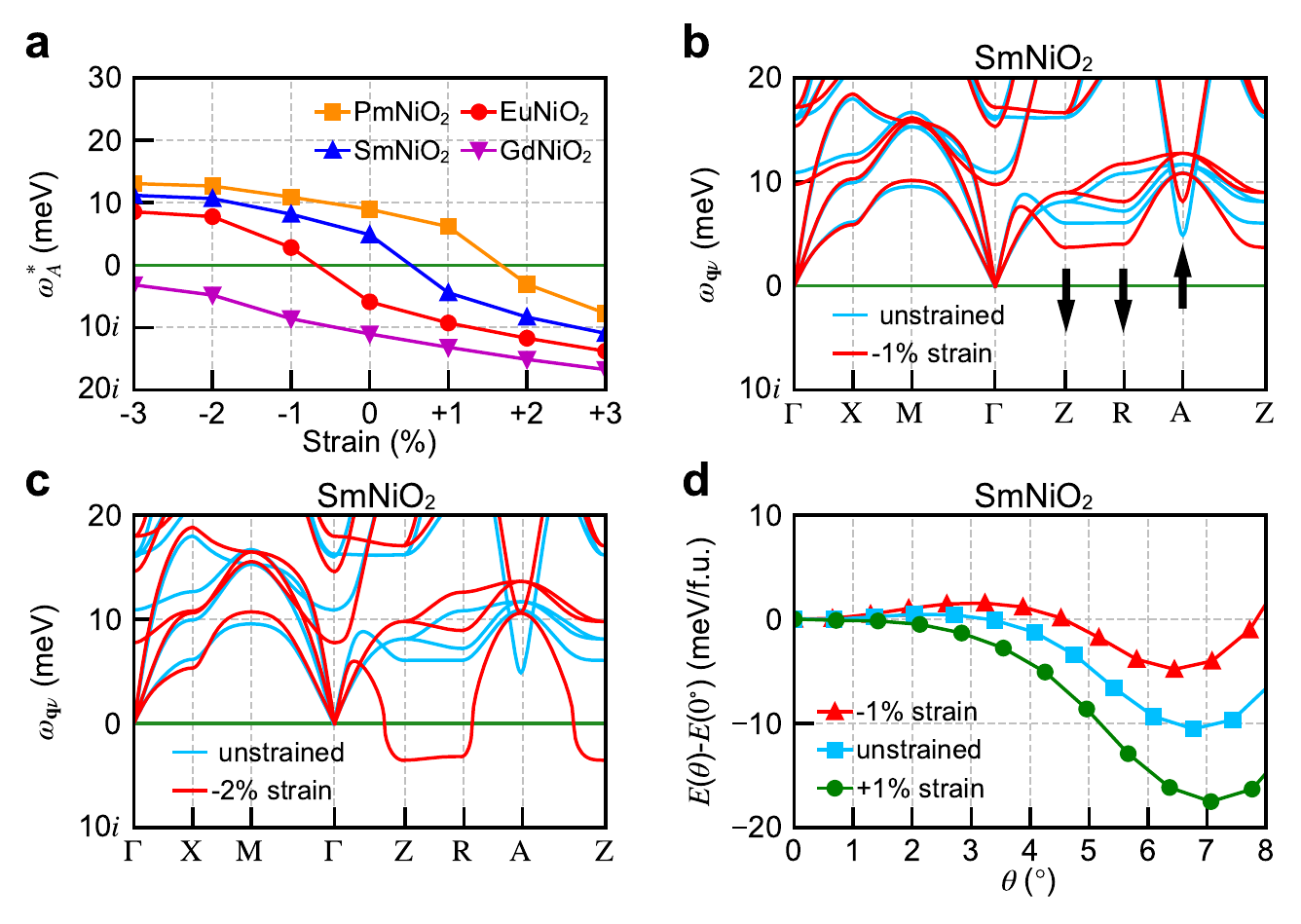}
\caption{\label{fig:strain}
\textbf{a}: The lowest phonon frequency at 
$A=(\pi,\pi,\pi)$ point 
of infinite-layer $R$NiO$_2$ in the $P4/mmm$ structure 
($R$ = Pm, Sm, Eu and Gd) as a function 
of biaxial strain. Negative (positive) strain means 
compressive (tensile) biaxial strain.
\textbf{b} and \textbf{c} : The phonon dispersions of 
infinite-layer SmNiO$_2$ in the 
$P4/mmm$ structure. The cyan curves are the phonon 
dispersions of fully-relaxed SmNiO$_2$ (i.e. without strain). 
The red 
curves in \textbf{b} and
\textbf{c} are the phonon dispersions of 
SmNiO$_2$ under 1\% and 2\% compressive strains, 
respectively. 
The black arrows highlight that with 
compressive strain, the 
lowest phonon mode at $A=(\pi,\pi,\pi)$ point is 
``hardened'', while the 
lowest phonon modes at $Z=(0,0,\pi)$ and $R=(\pi,0,\pi)$ 
are ``softened''.
\textbf{d}: The energy evolution of 
infinite-layer SmNiO$_2$ as a function of 
``NiO$_4$ square"
rotation angle $\theta$. The red, 
cyan and green 
symbols correspond to 1\% compressive strain, 
no strain and 
1\% tensile strain, respectively. 
For each case, the 
$P4/mmm$ structure is used as the energy reference.}
\end{figure}

Before we carefully compare the physical properties of $R$NiO$_2$ 
between the $I4/mcm$ structure and the $P4/mmm$ structure, we 
study epitaxial strain effects first. That is 
because superconductivity in infinite-layer nickelates is
observed in thin films rather than in 
bulk~\cite{Li2019, Li2020,Li2020a, Zeng2020, Osada2020,
Osada2020a,Ren2021, Osada2021, Zeng2021,He2021}. 
We investigate how epitaxial strain  
influences the phonon dispersion of infinite-layer $R$NiO$_2$, 
in particular, whether it may remove the 
imaginary phonon mode at $A$ point and 
thus stabilize the $P4/mmm$ structure. Experimentally, 
oxide thin films are grown along the $z$ axis with a 
biaxial strain imposed by substrates 
in the $xy$ plane. The biaxial strain (either compressive 
or tensile) typically ranges within 3\%~\cite{Schlom2008}.

Panel \textbf{a} of Fig.~\ref{fig:strain} shows the lowest phonon 
frequency at $A$ point $\omega_A^{*}$ of a few 
infinite-layer $R$NiO$_2$ in the $P4/mmm$ structure 
as a function of biaxial strain $\xi$. We select 
$R$ = Pm, Sm, Eu and Gd, which are close to the 
phase boundary where $\omega_A^{*}$ becomes imaginary 
(the complete phonon dispersions of those four nickelates under
epitaxial strain are found in Supplementary Materials~\cite{SI} Sec.~III). 
The biaxial strain is defined as 
$\xi = (a_{\textrm{sub}}-a)/a \times 100\%$, 
where $a_{\textrm{sub}}$ is the theoretical 
substrate lattice constant 
and $a$ is the DFT optimized 
lattice constant of infinite-layer $R$NiO$_2$ in the 
$P4/mmm$ structure. For each infinite-layer $R$NiO$_2$, 
we vary the strain $\xi$ and find that $\omega_A^{*}$ 
decreases with tensile strain and increases with 
compressive strain. 
However, we note that for infinite-layer nickelate
GdNiO$_2$, a compressive strain up 
to 3$\%$ can not remove the ``NiO$_4$ rotation" 
instability in 
the $P4/mmm$ structure. This is also true for other  
infinite-layer nickelates $R$NiO$_2$ with late 
lanthanide elements ($R$ = Gd-Lu). 
More importantly, we find that while compressive strain 
helps remove the phonon instability at $A$ point 
in the $P4/mmm$ structure, it may induce other 
phonon instabilities. Panel \textbf{b} compares the 
phonon dispersions of infinite-layer nickelate SmNiO$_2$ 
in the $P4/mmm$ structure under 1\% compressive strain 
versus without epitaxial strain. It shows that 
compressive strain ``hardens" the lowest phonon
frequency at $A$ point but ``softens" the lowest phonon 
frequencies at $Z$ and $R$ points. Under a compressive 
strain of 2\% or larger (see panel \textbf{c} and 
Fig. S3 in the Supplementary Materials~\cite{SI}), the lowest phonon 
frequencies at $Z$ and $R$ points become imaginary in 
the $P4/mmm$ structure. To summarize, for 
infinite-layer nickelates, tensile strain 
increases the phonon instability at $A$ point 
in the $P4/mmm$ structure; small compressive strain 
helps remove the phonon instability at $A$ point 
but larger compressive strain can cause 
other phonon instabilities at $Z$ and $R$ points.
Hence, epitaxial strain alone cannot 
substantially increase the 
stability of the $P4/mmm$ structure in infinite-layer 
nickelates $R$NiO$_2$. 

On the other hand, we find that for SmNiO$_2$, 
epitaxial strain can tune its energetics and 
structural properties.
Panel \textbf{d} of Fig.~\ref{fig:strain} shows the 
energy evolution of SmNiO$_2$ as 
a function of ``NiO$_4$ square" rotation angle $\theta$. 
We compare three different epitaxial strains: 1\% 
compressive (-1\%), 
no strain (0\%) and 1\% tensile (+1\%). From 1\% 
compressive 
strain to 1\% tensile strain, the energy difference 
between the $I4/mcm$ structure and the $P4/mcm$ structure 
monotonically increases from 4.8 to 17.5 meV/f.u. in its magnitude 
(indicating that the $I4/mcm$ structure gradually becomes 
more stable than the $P4/mmm$ structure). At the same time, 
the energy 
barrier from the $P4/mmm$ structure to the $I4/mcm$ structure 
decreases from 1.5 meV/f.u. (1\% compressive strain) 
to 0.5 meV/f.u. (no strain) and disappears (1\% tensile strain). 
The disappearance of the energy barrier indicates that under 1\% 
tensile strain, the $P4/mmm$ structure is no longer a local minimum 
in SmNiO$_2$ and it spontaneously transitions into the 
$I4/mcm$ structure.
Furthermore, the equilibrium ``NiO$_4$ square" rotation 
angle $\theta$ in the $I4/mcm$ structure also increases 
from $6.4^{\circ}$ to $7.1^{\circ}$ when the epitaxial 
strain changes from 1\% compressive to 1\% tensile strain. 
This shows that epitaxial strain can be 
used as a fine-tuning 
knob to delicately control the structural stability of 
infinite-layer SmNiO$_2$.

In the next three sections, we will study the entire 
lanthanide series of infinite-layer 
nickelates in the $I4/mcm$ structure and in the $P4/mcm$ 
structure. We compare the trends in electronic properties 
and magnetic properties between the two crystal structures.
For demonstration, we use SmNiO$_2$ as a prototype.

\subsection{\textit{P4/mmm} versus \textit{I4/mcm} structures: 
structural properties}

\begin{figure}[t]
\includegraphics[width=0.8\textwidth]{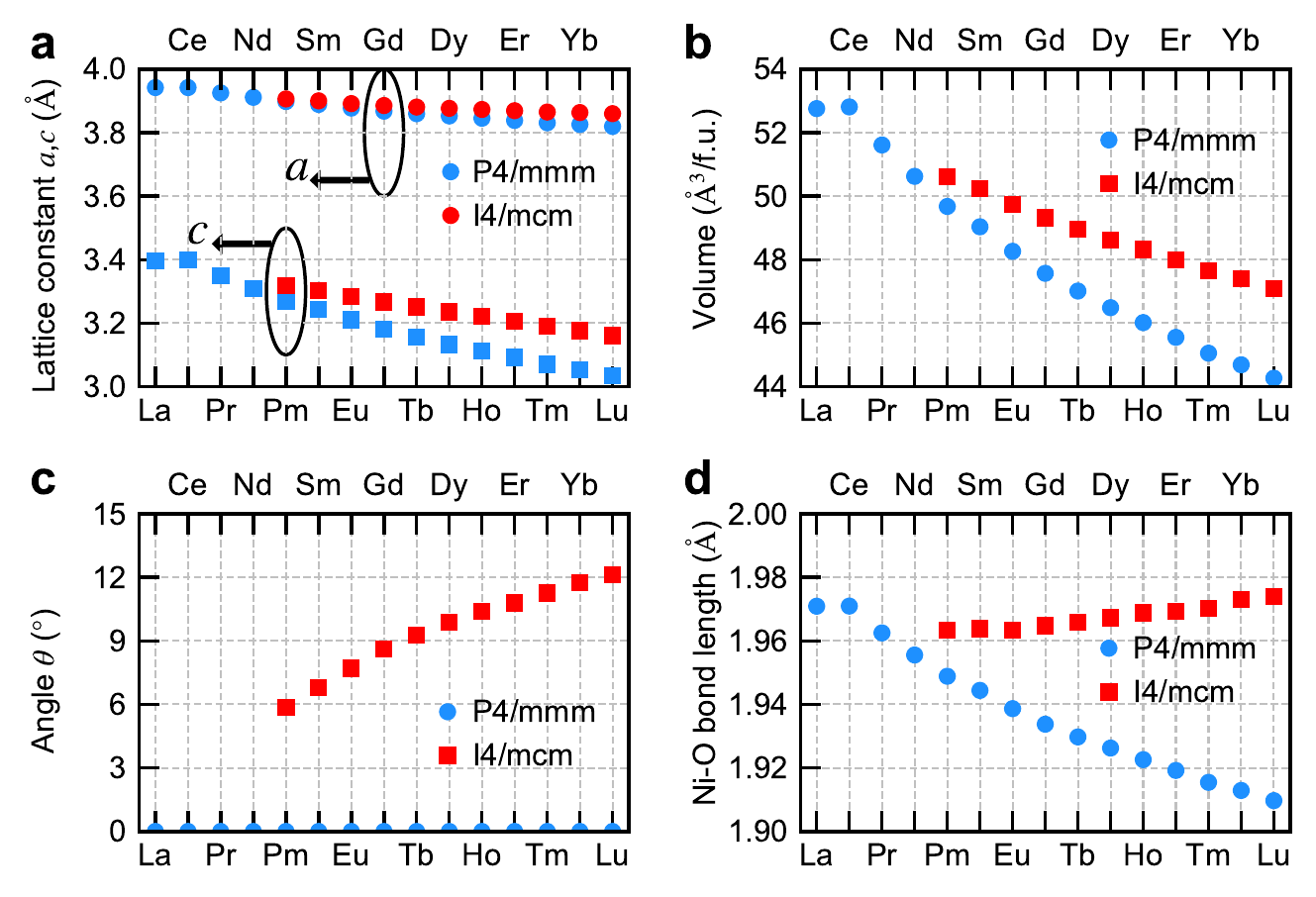}
\caption{\label{fig:lattice} Crystal information of infinite-layer 
nickelates $R$NiO$_2$ in the $I4/mcm$ structure (red symbols) 
and in the $P4/mmm$ 
structure (blue symbols). 
\textbf{a}: The lattice constants $a$ (circle symbols)
and $c$ (square symbols). 
For ease of comparison, we convert the lattice constants of 
the $I4/mcm$ structure into the pseudo-tetragonal lattice
constants $a$ and $c$.
\textbf{b}: Volume per $R$NiO$_2$ formula (f.u.).
\textbf{c}: ``NiO$_4$ square" rotation angle $\theta$. 
In the $P4/mmm$ structure, $\theta = 0^{\circ}$.
\textbf{d}: Ni-O bond length.}
\end{figure}

Fig.~\ref{fig:lattice} compares the structural properties of $R$NiO$_2$
between the $I4/mcm$ structure and the $P4/mmm$ structure. For infinite-layer 
nickelates in the first category ($R$ = La-Nd), we only study 
the $P4/mmm$ structure because the $I4/mcm$ structure cannot 
be stabilized in those nickelates.
Panels \textbf{a} shows the lattice constants $a$ and $c$ of 
the $I4/mcm$ and $P4/mmm$ structures. For ease of comparison 
to the $P4/mmm$ structure, we convert the lattice constants of 
the $I4/mcm$ structure into the pseudo-tetragonal lattice constants
$a$ and $c$ (see Supplementary Materials~\cite{SI} Sec.~I). The general 
trend is similar in the two structures that $a$ and $c$ get smaller when $R$ traverses the 
lanthanide series~\cite{Been2021}. 
For a given $R$, the $a$ and $c$ lattice constants are larger 
in the $I4/mcm$ structure than in the $P4/mmm$ structure. 
Panel \textbf{b} shows the volume per Ni atom of the $I4/mcm$ 
and $P4/mmm$ structures. Consistent with the trends of the 
lattice constants, the volume decreases as $R$ traverses the 
lanthanide series. The key difference between the $I4/mcm$ structure and the 
$P4/mmm$ structure lies in the ``NiO$_4$ square" rotation.
In panel \textbf{c}, we show the ``NiO$_4$ square" 
rotation angle $\theta$. In the $P4/mmm$ structure, $\theta=0$ by 
definition. We find that $\theta$ increases in the $I4/mcm$ structure, 
as we traverse from Pm to Lu. A finite $\theta$ 
means that the in-plane Ni-O-Ni bond angle is reduced from 
the ideal 180$^{\circ}$. A direct consequence of $\theta$ is the 
elongation of Ni-O bond. In the $P4/mmm$ structure, the Ni-O 
bond length is simply half of the lattice constant $a$, which 
decreases as $R$ traverses the lanthanide series. By contrast, in the $I4/mcm$ 
structure, Ni-O bond length is elongated compared to that in the $P4/mmm$ 
structure and it slowly increases as $R$ traverses the lanthanide series.
We note that the ``NiO$_4$ square" rotation $\theta$ and the volume reduction are 
two competing forces on the Ni-O bond length. The former, which is absent in the 
$P4/mmm$ structure, is more dominating in the $I4/mcm$ structure.  
A similar picture of these two competing forces is also found in LiNbO$_3$ under 
hydrostatic pressure~\cite{Xia2021}.
The different behaviors of Ni-O-Ni bond angle and Ni-O bond length in the $I4/mcm$ structure 
versus in the $P4/mmm$ structure will have important influences on the electronic 
properties of $R$NiO$_2$.

\subsection{\textit{P4/mmm} versus \textit{I4/mcm} structures: 
electronic properties}

\begin{figure}[t]
\includegraphics[width=0.8\textwidth]{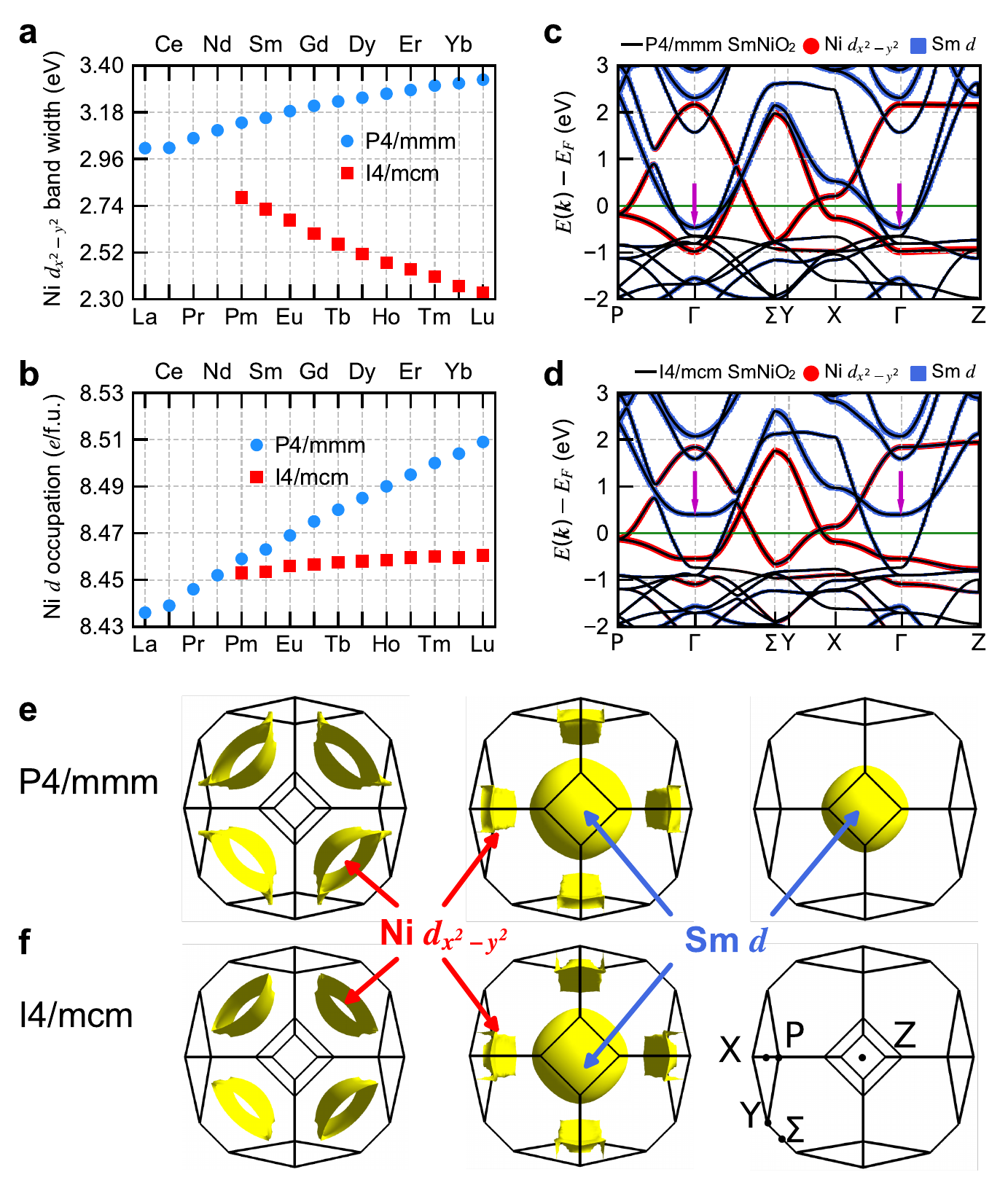}
\caption{\label{fig:electron} Electronic properties of 
infinite-layer nickelates $R$NiO$_2$ in the $I4/mcm$ structure 
and in the $P4/mmm$ structure. \textbf{a} and \textbf{b}: bandwidth 
of Ni $d_{x^2-y^2}$ orbital and Ni $d$ occupancy.
The red (blue) symbols refer to the
$I4/mcm$ structure (the $P4/mmm$ structure).
\textbf{c} and \textbf{d}: Electronic structure of 
SmNiO$_2$ in the $P4/mmm$ structure and in the $I4/mcm$ 
structure, respectively. The red (blue) symbols highlight 
the atomic projection onto the Ni $d_{x^2-y^2}$ orbital 
(Sm $d$ orbitals). 
The purple arrows in \textbf{c} and \textbf{d} highlight
that a Sm-$d$-derived band is removed from 
the Fermi level and is pushed up to higher energy 
via the ``NiO$_4$ square"  rotation 
in the $I4/mcm$ structure.
\textbf{e} and \textbf{f}: 
Fermi surface of SmNiO$_2$ in the $P4/mmm$ structure and 
in the $I4/mcm$ structure, respectively. The red (blue) arrows
highlight the Fermi surface sheets that are composed of 
Ni $d_{x^2-y^2}$ orbital (Sm $d$ orbitals).}
\end{figure}

Panel \textbf{a} of Fig.~\ref{fig:electron} compares the 
Ni $d_{x^2-y^2}$ bandwidth of 
the $I4/mcm$ structure and the $P4/mmm$ structure. 
Consistent with the previous studies~\cite{Kapeghian2020,Been2021}, 
the Ni $d_{x^2-y^2}$ bandwidth 
of the $P4/mmm$ structure 
monotonically increases when $R$ traverses the lanthanide 
series. That is because the Ni-O bond length of the $P4/mmm$ 
structure monotonically 
decreases, which increases the Ni-O hopping and thus 
the bandwidth. By contrast, 
we find that the Ni $d_{x^2-y^2}$ 
bandwidth of the $I4/mcm$ structure monotonically 
\textit{decreases} when $R$ traverses 
the lanthanide series. That is consistent with the trend of
a decreasing Ni-O-Ni bond angle and an increasing
Ni-O bond length, both of which reduce the overlap between
Ni-$d_{x^2-y^2}$ and O-$p$ orbitals and thus suppress the Ni-O hopping~\cite{Kumah2014}.
More importantly, for a given lanthanide element $R$, 
the Ni $d_{x^2-y^2}$ bandwidth 
of the $I4/mcm$ structure is substantially smaller than that of 
the $P4/mmm$ structure. For example, for SmNiO$_2$, its 
Ni $d_{x^2-y^2}$ bandwidth is 3.2 eV in the $P4/mmm$ structure and 
is reduced to 2.7 eV in the $I4/mcm$ structure. This indicates that 
for the same value of $U$ on Ni $d_{x^2-y^2}$ orbital, 
correlation strength is increased in the $I4/mcm$ structure, 
compared to the $P4/mmm$ structure. In addition, we 
also compare the Ni $d$ orbital occupancy $N_d$ between the 
$I4/mcm$ structure and the 
$P4/mmm$ structure in panel \textbf{b}. 
Ref.~\cite{Wang2012} shows that the metal $d$ orbital 
occupancy $N_d$ is a good measure 
of $p$-$d$ hybridization in complex oxides.
We find that as $R$ traverses the lanthanide 
series, $R$NiO$_2$ in the $P4/mmm$ structure 
has a progressively increased $N_d$. 
This is consistent with 
the previous study~\cite{Been2021} which shows that 
the O-$p$ content 
monotonically decreases across the lanthanide series.
However, for $R$NiO$_2$ in the $I4/mcm$ structure, 
$N_d$ almost stays a constant as $R$ traverses the 
lanthanide series. By analyzing the density of states of $R$NiO$_2$
(see Supplementary Materials~\cite{SI} Sec.~VI), 
we find that in the $P4/mmm$ structure, running across 
the lanthanide series, the centroid of O-$p$ states monotonically 
decreases to lower energy, by about 1 eV from La to Lu~\cite{Been2021}. 
This change increases the charge-transfer energy and 
decreases the $p$-$d$ hybridization. But in the 
$I4/mcm$ structure, the ``NiO$_4$ square" rotation 
counteracts this effect and the 
centroid of O-$p$ states almost does not move
across the lanthanide series.  
As a result, the Ni $d$ occupancy $N_d$ and $p$-$d$ hybridization 
change marginally across the lanthanide 
series of $R$NiO$_2$. We note that for a 
given lanthanide element $R$, $N_d$ is smaller in the $I4/mcm$ 
structure than in the $P4/mmm$ structure. A smaller $N_d$ 
corresponds to a smaller critical $U$ value 
for the metal-insulator transition~\cite{Wang2012}
, i.e. $R$NiO$_2$ in the $I4/mcm$ structure is closer to the Mott 
insulating phase than that in the $P4/mmm$ structure. 

Next we study the electronic band structure and Fermi surface of SmNiO$_2$ as a prototype 
(very similar results are also obtained in infinite-layer 
nickelates close to the phase boundary $R$NiO$_2$ with $R$ = 
Pm, Eu and Gd, see Supplementary Materials~\cite{SI} Sec.~V). 
Panel \textbf{c} of Fig.~\ref{fig:electron} shows 
the band structure of 
SmNiO$_2$ in the $P4/mmm$ structure. 
Due to the cell-doubling, there are four bands 
that cross the Fermi level: two 
are Ni-$d_{x^2-y^2}$-derived bands and 
the other two are Sm-$d$-derived bands.
In the unfolded Brillouin zone (BZ), the 
Sm-$d$-derived band crosses the Fermi level and 
results in two 
electron pockets: one is at $\Gamma$ point 
and the other is at $A$ point. After band 
folding, in the body-centered-tetragonal 
Brillouin zone (BCT-BZ)~\cite{Been2021}, the electron pocket 
that is originally at $A$
point in the unfolded BZ is mapped to $\Gamma$ point, 
leading to 
two electron pockets at $\Gamma$ point. This is 
clearly seen in panel \textbf{e}, which shows the 
Fermi surface of SmNiO$_2$ in the $P4/mmm$ structure.
These $\Gamma$-centered 
electron pockets are one of the main differences 
between 
infinite-layer nickelates and superconducting cuprates and 
their role is still under debate~\cite{Jiang2019,Botana2020,Karp2020a,
Adhikary2020,Zhang2020d,Nomura2019,Gu2020,Hirayama2020}. 
Panel \textbf{d} of Fig.~\ref{fig:electron} shows 
the band structure of 
SmNiO$_2$ in the $I4/mcm$ structure. 
Compared to the $P4/mmm$ 
structure, the ``NiO$_4$ square" rotation removes 
one Sm-$d$-derived band (highlighted by the purple arrows) 
away from the Fermi surface and pushes it to 
higher energy. In the corresponding Fermi surface 
(panel \textbf{f}), one $\Gamma$-centered  
electron pocket vanishes. This is an interesting 
result in that 1) compared to the 
$P4/mmm$ structure, the Fermi surface 
of SmNiO$_2$ in the $I4/mcm$ structure more closely 
resembles that of CaCuO$_2$; and 2) the ``NiO$_4$ square" rotation 
in the $I4/mcm$ structure effectively acts as 
hole doping in SmNiO$_2$. To demonstrate the second point 
more clearly, we calculate the band structure and 
Fermi surface of Sr$_{0.2}$Sm$_{0.8}$NiO$_2$ in the $P4/mmm$ 
structure (see the Supplementary Materials~\cite{SI} Sec.~VII) and we find that 
they are similar to pristine SmNiO$_2$ in the $I4/mcm$ structure. 

\subsection{\textit{P4/mmm} versus \textit{I4/mcm} structures: 
magnetic properties}

Next we study the magnetic properties of $R$NiO$_2$ and compare 
the $I4/mcm$ structure and the $P4/mmm$ structure.
We use the charge-only DFT+$U$+$J$ 
method~\cite{Park2015,Chen2015,Chen2016} in which the spin 
polarization is broken by the $U/J$ extension rather than the 
spin-dependent exchange-correlation functional. The advantage of 
using this method is that when $U/J$ parameters approach zero, 
we will recover our non-spin-polarized DFT results. 
We first fix the optimized crystal structure that is obtained from the 
non-spin-polarized (nsp) DFT calculations, upon which 
the electronic structure calculations are performed. 
This is the convention of some previous DFT+$U$ and DFT+dynamical 
mean field theory (DFT+DMFT) studies~\cite{Been2021,Ryee2020,Choi2020,Liu2020a,Choi2020a,Lechermann2020,Lechermann2020a,Leonov2020a,Petocchi2020,Lechermann2021}. Then we relax the crystal structure within the charge-only DF+$U$+$J$ method and discuss the 
relaxation effects on magnetic properties. 

\begin{figure}[t]
\includegraphics[width=0.8\textwidth]{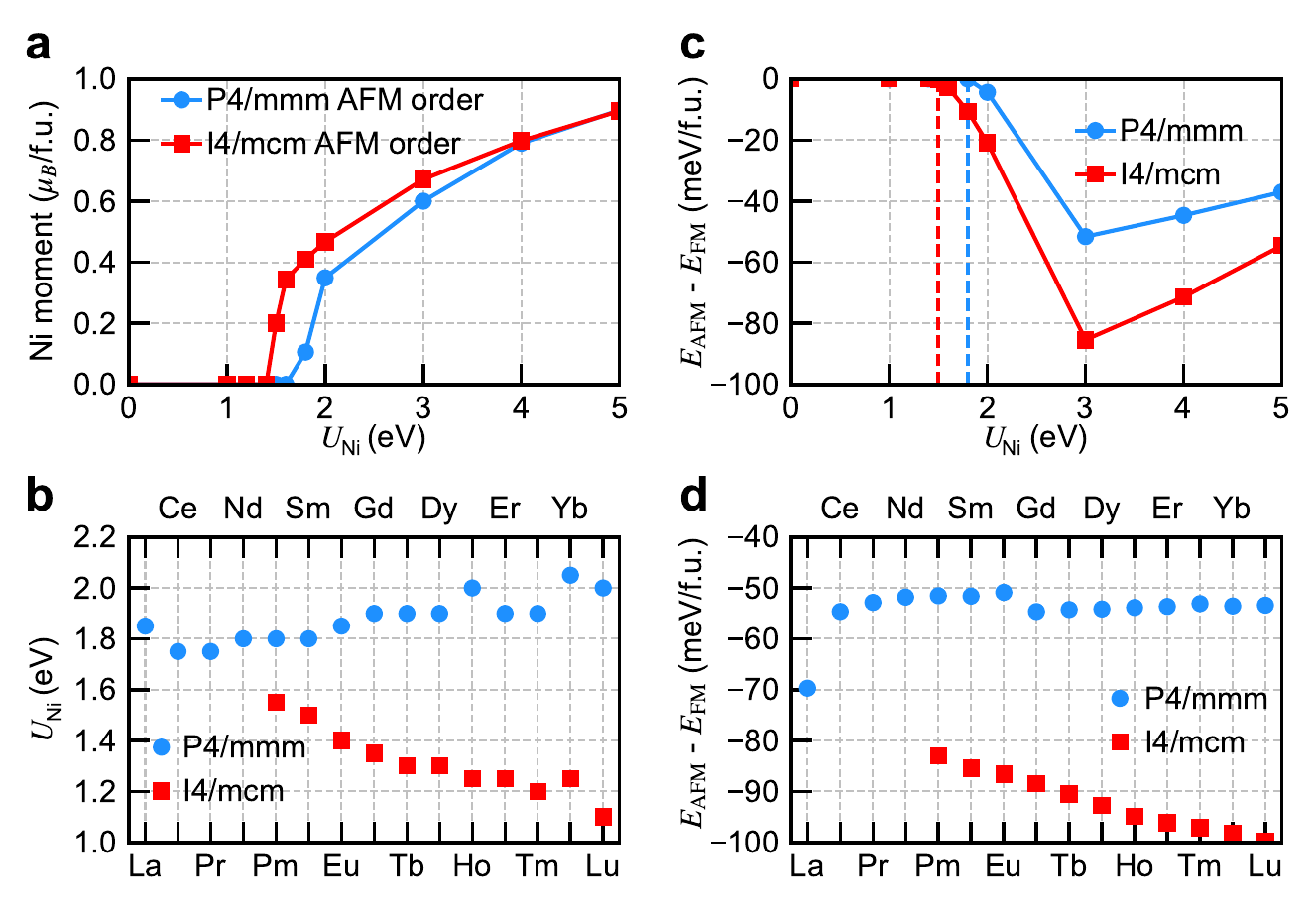}
\caption{\label{fig:mag} Magnetic properties of 
infinite-layer nickelates $R$NiO$_2$ in the $I4/mcm$ 
structure (red symbols) and in the $P4/mmm$ structure 
(blue symbols). 
\textbf{a}: The magnetic moment on Ni atom of SmNiO$_2$ 
in rocksalt antiferromagnetic ordering (AFM) as a function 
of $U_{\textrm{Ni}}$.
\textbf{b}: The
critical $U_\mathrm{Ni}$ for 
rocksalt antiferromagnetic ordering 
across the lanthanide series 
of infinite-layer nickelates $R$NiO$_2$.
\textbf{c}: The energy difference 
between rocksalt antiferromagnetic 
ordering and ferromagnetic ordering
of SmNiO$_2$ as a function of $U_\mathrm{Ni}$. 
The dashed lines highlight the critical
$U_\mathrm{Ni}$ for rocksalt antiferromagnetic 
ordering.
\textbf{d}: The energy 
difference between rocksalt antiferromagnetic 
ordering and ferromagnetic ordering at 
$U_\mathrm{Ni}$ = 3 eV across the lanthanide series 
of infinite-layer nickelates $R$NiO$_2$.}
\end{figure}

We first study rocksalt antiferromagnetic ordering 
(ordering wave vector $\textbf{q}=(\pi,\pi,\pi)$) in
SmNiO$_2$~\cite{Liu2020a}. 
Panel \textbf{a} of Fig.~\ref{fig:mag} shows the magnetic 
moment on Ni atom as a function of $U_{\textrm{Ni}}$ 
(throughout the calculations, we set 
$J_{\textrm{Ni}}= 0.15 U_{\textrm{Ni}}$ ~\cite{Giovannetti2014}). 
We compare 
the Ni magnetic moment between the $I4/mcm$ structure 
and the $P4/mmm$ structure. We find that within the charge-only DFT+$U$+$J$ method, 
the critical effective $U_{\textrm{Ni}}$ for rocksalt antiferromagnetic 
ordering is reduced from 1.8 eV in the $P4/mmm$ structure 
to 1.5 eV in the $I4/mcm$ structure. This is consistent 
with the bandwidth reduction effect that correlation strength 
is increased on Ni $d_{x^2-y^2}$ orbital 
in the $I4/mcm$ structure, which favors the 
formation of long-range magnetic ordering. 
Next in panel \textbf{b}, we study 
the entire lanthanide series of infinite-layer nickelates 
$R$NiO$_2$ and compare the 
critical $U_{\textrm{Ni}}$ for the $I4/mcm$ 
structure and for the $P4/mmm$ structure. 
We find that 
the critical $U_{\textrm{Ni}}$ for the $I4/mcm$ structure 
is always smaller than that for the $P4/mmm$ structure, and 
from PmNiO$_2$ to LuNiO$_2$, the reduction in the critical 
$U_{\textrm{Ni}}$ becomes more substantial. This 
feature is consistent with the trend of Ni $d_{x^2-y^2}$ 
bandwidth reduction 
(see Fig.~\ref{fig:electron}\textbf{a}). 
In addition, we study the energy difference between 
rocksalt antiferromagnetic ordering and 
ferromagnetic ordering 
$\Delta E = E_{\textrm{AFM}} - E_{\textrm{FM}}$ 
as a function of $U_{\textrm{Ni}}$. Panel \textbf{c} shows 
that for SmNiO$_2$, both in 
the $I4/mcm$ structure and in the $P4/mmm$ 
structure, when $U_{\textrm{Ni}}$ exceeds the critical 
value (highlighted by the two dashed lines), 
rocksalt antiferromagnetic 
ordering has lower energy than ferromagnetic 
ordering ($\Delta E$ is negative). 
However, the magnitude of $|\Delta E|$ is larger 
in the $I4/mcm$ structure than in the $P4/mmm$ structure 
when the long-range magnetic order is stabilized in SmNiO$_2$.
The results in panel \textbf{c} 
indicate that given the same value of $U_{\textrm{Ni}}$, 
the ``NiO$_4$ square" rotation in the $I4/mcm$ structure further stabilizes the 
rocksalt antiferromagnetic ordering over the ferromagnetic ordering.
We repeat the same calculations for the entire lanthanide 
series of infinite-layer nickelates $R$NiO$_2$ and show in 
panel \textbf{d} the energy difference $\Delta E$ at 
$U_{\textrm{Ni}} = 3$ eV. We find that $\Delta E$ is 
negative and its magnitude is larger in the 
$I4/mcm$ structure than in the $P4/mmm$ structure
for the entire series of $R$NiO$_2$ ($R$ = Pm-Lu).

Next within the  
charge-only DFT+$U$+$J$ method, we relax the 
crystal structure of SmNiO$_2$ for each given $U$ and $J$ 
(see Supplementary Materials~\cite{SI} Sec.~VIII). 
We find that adding $U$ and $J$ terms does not 
considerably change the optimized lattice constants 
and the ``NiO$_4$ square" rotation angle. 
However, it is noted that a weak ``cusp" feature emerges 
at the critical $U_{\textrm{Ni}}$ 
when the long-range magnetic ordering is stabilized. 
Using the optimized crystal structure from the charge-only 
DFT+$U$+$J$ method, we still find that 1) the 
critical $U_{\textrm{Ni}}$ for the $I4/mcm$ structure 
is smaller than that for the $P4/mmm$ structure, 2) 
rocksalt-antiferromagnetic ordering is more stable 
than ferromagnetic ordering in both crystal structures, and 
3) the magnitude of the energy difference 
between rocksalt-antiferromagnetic ordering and 
ferromagnetic ordering $|\Delta E|$ is larger in the $I4/mcm$ structure 
than in the $P4/mmm$ structure. All these results 
are qualitatively consistent with the previous ones that are obtained 
by using the nsp-DFT optimized crystal structure.

\section{Conclusion}
In conclusion, we perform first-principles calculations 
to study structural, electronic and magnetic properties 
of the entire lanthanide series of infinite-layer 
nickelates $R$NiO$_2$.
We find that the widely-studied 
$P4/mmm$ structure is only dynamically stable 
when $R$ is an early lanthanide element (La-Sm). 
For late lanthanide elements (Eu-Lu), 
an unstable phonon mode appears at $A$ point 
in the $P4/mmm$ structure, which
corresponds to an out-of-phase 
``NiO$_4$ square" rotation about the $z$ axis. 
For infinite-layer nickelates with late 
lanthanide elements, 
condensation of this phonon mode in the $P4/mmm$
structure lowers the total 
energy and leads to a new $I4/mcm$ crystal structure.
Special attention is paid to 
two borderline cases PmNiO$_2$ and SmNiO$_2$, in 
which both the $P4/mmm$ structure 
and the $I4/mcm$ structure are local minimums. 
When epitaxial strain is imposed on infinite-layer 
nickelates, tensile strain further increases 
the dynamical instability at $A$ point in the 
$P4/mmm$ structure, while compressive strain 
``hardens'' the phonon at $A$ point but ``softens'' the 
phonons at $Z$ and $R$ points in the $P4/mmm$ structure.
Furthermore, epitaxial strain can fine-tune the 
energy difference between the $I4/mcm$ structure 
and the $P4/mmm$ structure when they are both 
dynamically stable in $R$NiO$_2$ (such as SmNiO$_2$). 

We use the new $I4/mcm$ crystal structure 
to study the trends of electronic and magnetic 
properties of $R$NiO$_2$. We find that compared to 
the $P4/mmm$ structure, the Ni $d_{x^2-y^2}$ 
bandwidth of $R$NiO$_2$ is substantially 
reduced in the $I4/mcm$ structure, 
which implies an increased correlation 
strength in the new $I4/mcm$ structure.
In addition, the Ni $d$ occupancy of 
$R$NiO$_2$ gets smaller 
in the $I4/mcm$ structure than that 
in the $P4/mmm$ structure, 
which means a small critical $U_{\textrm{Ni}}$ for the 
metal-insulator transition~\cite{Wang2012}.
Furthermore, the electronic 
structure and Fermi surface of $R$NiO$_2$ 
become ``cleaner" in the $I4/mcm$ structure 
than in the $P4/mmm$ structure, 
because one lanthanide-$d$-derived band 
is removed from the Fermi level and thus a lanthanide-$d$-derived 
electron pocket disappears at $\Gamma$ point.
Finally, the critical $U_{\textrm{Ni}}$ to stabilize the 
rocksalt antiferromagnetic ordering in $R$NiO$_2$ 
is reduced from the $P4/mmm$ structure to the 
$I4/mcm$ structure. 
All these results imply that correlation effects are enhanced and  
Mott physics plays a more important role 
in the new $I4/mcm$ crystal structure of infinite-layer 
$R$NiO$_2$. Hence, if $R$NiO$_2$ in the 
$I4/mcm$ crystal structure 
can be synthesized in experiment, it will provide a 
closer analogy to infinite-layer cuprate CaCuO$_2$.
Our work suggests that 
among the lanthanide series of infinite-layer nickelates,
the most promising candidate to crystallize in the $I4/mcm$ 
structure is SmNiO$_2$.

We finally note that structure-property 
relations have been widely studied in 
complex oxides, such as perovskite nickelates 
and manganites~\cite{Medarde1997,Salamon2001,doi:10.1080/01411590801992463,Middey2016,Catalano2018}.
Spontaneous structural distortions, 
such as Jahn-Teller, breathing, rotations and 
tilts of oxygen octahedra, turn out to have 
substantial impacts on the physical properties of complex oxides
~\cite{Rondinelli2012,Benedek2013,Zhai2014,Liao2016,Chen2018}.
Our work reveals a similar coupling between crystal structure and electronic structure
in infinite-layer nickelates $R$NiO$_2$: by substituting the 
lanthanide element $R$, we can control the rotation of 
``NiO$_4$ square", which tunes the underlying electronic 
structure and may potentially favor superconductivity.

\textit{Note added}: after the completion of our work, we became aware of
Refs.~\cite{Bernardini2021arxiv, Alvarez2021, Zhang2022}, which also study
structural distortions in infinite-layer nickelates $R$NiO$_2$ and YNiO$_2$.
The authors of Ref.~\cite{Bernardini2021arxiv} find that YNiO$_2$ is also prone
to the ``NiO$_4$ square" rotation. The authors of Ref.~\cite{Alvarez2021}
show that due to the $R$-to-Ni
cation mismatch, the ``NiO$_4$ square" is prone to various types of
rotations. For $R$ of a small ionic radius such as Y, the ground state
structure has the orthorhombic $Pbnm$ symmetry with the $a^-a^-c^+$
rotation pattern. For $R$ of a moderate ionic radius such as Gd, the
ground state structure has the tetragonal $I4/mcm$ symmetry with the
$a^0a^0c^-$ rotation, which is consistent with our results. The authors of Ref.~\cite{Zhang2022}
also find the dynamical instability of
``NiO$_4$ square" rotation in infinite-layer nickelates $R$NiO$_2$.
In addition, they study the electronic and magnetic
properties of $R$NiO$_2$ at finite temperatures.

\section*{Acknowledgement}
We are grateful to Andrew Millis for useful discussions.

\newpage
\clearpage


\end{document}